\documentclass[conference]{IEEEtran}
\IEEEoverridecommandlockouts
\usepackage{cite}
\usepackage{amsmath,amssymb,amsfonts}
\usepackage{algorithmic}
\usepackage{mdframed}
\usepackage{framed}
\usepackage{graphicx}
\usepackage{textcomp}
\usepackage{xcolor}
\usepackage{enumitem}
\usepackage{mathtools}
\usepackage{comment}
\usepackage{float}
\DeclarePairedDelimiter\floor{\lfloor}{\rfloor}

\newmdtheoremenv{theorem}{Theorem}
\newmdtheoremenv[backgroundcolor=gray!20]{result}{Result}
\newcommand{\Mod}{\mathrm{mod}\ }
\usepackage{subfig}
\usepackage[ruled,vlined]{algorithm2e}
\usepackage{multirow}
\def\BibTeX{{\rm B\kern-.05em{\sc i\kern-.025em b}\kern-.08em
    T\kern-.1667em\lower.7ex\hbox{E}\kern-.125emX}}

\begin{document}

\title{Preventing Denial of Service Attacks in IoT Networks through Verifiable Delay Functions}
%\title{Preventing Denial of Service Attacks in Distributed Ledger Technologies for the Internet of Things}

\author{\IEEEauthorblockN{Vidal Attias}
\IEEEauthorblockA{\textit{IOTA Foundation}\\
Berlin, Germany\\
vidal.attias@iota.org}
\and
\IEEEauthorblockN{Luigi Vigneri}
\IEEEauthorblockA{\textit{IOTA Foundation}\\
Berlin, Germany\\
luigi.vigneri@iota.org}
\and
\IEEEauthorblockN{Vassil Dimitrov}
\IEEEauthorblockA{\textit{IOTA Foundation}\\
Berlin, Germany\\
vassil@iota.org}
}

\maketitle

\begin{abstract}
Permissionless distributed ledgers provide a promising approach to deal with the Internet of Things (IoT) paradigm. Since IoT devices mostly generate data transactions and micro payments, distributed ledgers that use fees to regulate the network access are not an optimal choice. In this paper, we study a feeless architecture developed by IOTA and designed specifically for the IoT. Due to the lack of fees, malicious nodes can exploit this feature to generate an unbounded number of transactions and perform denial of service attacks. We propose to mitigate these attacks through verifiable delay functions. These functions, which are non-parallelizable, hard to compute and easy to verify, have been formulated only recently. In our work, we design a denial of service prevention mechanism which addresses network heterogeneity, limited node computational capabilities and hardware-specific implementation optimizations. Verifiable delay functions have mostly been studied from a theoretical point of view, but little has been done in tangible applications. Hence, this paper can be considered as a pioneer work in the field, since it builds a bridge between this theoretical mathematical framework and a real-world problem.
\end{abstract}


\section{Introduction}\label{sec:introduction}

The \textit{Internet of Things} (IoT) paradigm has initiated a revolution in human lives through enhanced user experience and new applications~\cite{Pawar2016}. While IoT devices first reached consumers through small-scale proprietary networks~\cite{Ande2020}, nowadays these sensors can be embedded into mobile devices, industrial equipment, environmental sensors, medical devices, and more. The widespread availability and the increasing interest in IoT applications make necessary the creation of secure networks dealing with thousands or millions of IoT sensors.

%This can be exploited by hackers to generate huge networks of hacked devices used to run Denial of Service attacks on services. For example in 2018, the Mirai botnet gathered around 150.000 devices and launched a record attack against OVH reaching a peak of 1.7 TB per second \cite{Kolias2017}.

\subsection{Distributed ledger technologies for the IoT}

Due to the limited resources (e.g., CPU, storage) available, the design of secure and efficient protocols for IoT networks is challenging, and proprietary solutions are not anymore able to cope with their evolving structure. In this paper, we investigate the recent \textit{distributed ledger technology}\footnote{A DLT is a consensus of replicated, shared, and synchronized data without any central administrator or centralized data storage.} (DLT) paradigm as a secure way to deal with permissionless decentralized IoT networks. In the seminal work on \textit{blockchain}~\cite{Nakamoto}, specific network nodes (\textit{miners}) validate information through the solution of a cryptographic puzzle (\textit{Proof of Work}~\cite{Back2002}) in return for some monetary rewards (\textit{fees}). Apart from the obvious economical and environmental issues~\cite{Becker2013}, this protocol is not suitable for the IoT due to fees and to the high workload each node must perform. However, the interest on the topic has favored the proliferation of many alternative DLTs, some of them focusing on the idea of managing the IoT ecosystem. In particular, in this paper we consider the IoT-oriented permissionless DLT developed by IOTA~\cite{Popov2018}.

The IOTA protocol builds the \textit{Tangle}, a directed acyclic graph, where each vertex is a \textit{transaction}\footnote{A transaction is a message that transfers data or funds between nodes.}. Each time a user wants to issue a transaction, she has to verify and approve two recent transactions which then form the edges. This way, the integrity of the Tangle is ensured by the work of the users themselves rather than by a different economic set of nodes, like in the case of blockchain's miners. Furthermore, no fees are imposed by the protocol due to the lack of miners, enabling micro and data transactions, fundamentals in IoT networks.

Unlike blockchain where the mining layer imposes a filter on which transactions can be written into the ledger, the Tangle lacks an intrinsic access control algorithm. Since the concepts of miners and users are merged, IOTA is vulnerable to \textit{denial of service} (DoS) attacks, especially in IoT networks where resources are limited. A DoS attack is a method used to disrupt legitimate users' access to a target network or a certain resource. This is typically done by overloading the target with a massive amount of traffic. Traditionally, DoS attacks have been used to target banks, governments or online commercial retailers. However, since the cryptocurrency market represents nowadays hundreds of billions of US dollars~\cite{ElBahrawy2017}, DLTs have also become a popular target for DoS attacks. Hence, securing the stability of these networks is a crucial economical question.

\subsection{Challenges and contributions}

In this paper, we propose a lightweight access control mechanism for DLTs. The task is particularly challenging as our design must satisfy the following requirements: (i) The mechanism cannot involve fee-based spam prevention since IoT devices often need to send data transactions, rather than monetary payments; (ii) we target heterogeneous networks where \textit{any} node can join the network activity by issuing and writing transactions into the ledger independently of their capabilities; (iii) we consider the presence of malicious nodes trying to spam the network and perform DoS attacks.

Inspired by the seminal work in spam prevention by Dwork and Naor~\cite{Dwork}, and by the recently renewed interest in the field brought by Boneh et al.~\cite{Boneh2019}, we design our solution based on \textit{verifiable delay functions} (VDFs). VDFs are functions that require a preset number of iterations to complete. While VDFs and Proof of Work (PoW) both aim the evaluator to spend some time to compute a puzzle, the former is mainly based on sequentially computing a function (iterative squaring in RSA groups~\cite{Wesolowski, Pietrzak} or points addition in elliptic curves~\cite{DeFeo}) which \textit{cannot} be parallelized, making specialized hardware not able to substantially speed up the puzzle computation. We highlight that, unlike in PoW-based blockchains, the cryptographic puzzle discussed in this paper is only used to prevent DoS attacks, and does not affect the consensus. Our contributions are threefold:
\begin{itemize}[leftmargin=*]
    \item \textit{Theory.} We design a DoS prevention mechanism where nodes are required to compute exactly $\tau$ modular squarings of a given input message. Our mechanism is based on~\cite{Wesolowski} which presents characteristics in line with the IoT use case~\cite{Attias2020}, as we will show in the rest of the paper.
    \item \textit{Implementation.} We analyze the VDF evaluation time on different hardware (IoT, laptop, FPGA), and we optimize the time needed to verify the correctness of the puzzle through multiexponentiation techniques~\cite{Yen1994, Moller}.
    \item \textit{Analysis.} We perform extensive experimental evaluations to compare VDFs and PoW on different hardware. In particular, we will show that large monetary resources cannot help to speed up the VDF evaluation.
\end{itemize}

We would like to stress here that, to the best of our knowledge, this is one of \textit{first} studies which analyzes, optimizes and implements VDFs in a real-world problem. Furthermore, while we build our solution on top of the IOTA Tangle, we aim this work to be a source of inspiration for other IoT-oriented permissionless decentralized networks.

Finally, we mention that, in distributed systems, each identity-based protocol has to deal with the so-called \textit{Sybil attack}~\cite{Douceur2002}, where participants may create counterfeit identities in order to, e.g., have a larger weight in a voting protocol or overcome the access control mechanism. In the IOTA protocol, the proliferation of Sybil nodes is mitigated through the definition of a reputation system based on stake. The details of this approach are beyond the scope of this paper, and we defer the interested reader to~\cite{CoordicieTeam2019}.

% Outline

\subsection{Outline}
The rest of the paper is organized as follows: First, we present some background information about VDFs in Section~\ref{sec:vdf}; then, in Section~\ref{sec:model} we introduce the system model and the problem statement; after that, we propose our VDF-based DoS prevention mechanism and we provide its computational complexity analysis in Section~\ref{sec:protocol}; finally, we validate our findings through simulations in Section~\ref{sec:simulation}, and we conclude our paper in Section~\ref{sec:conclusion}.
\section{Verifiable Delay Functions}\label{sec:vdf}

DoS prevention is a field where functions similar to VDFs have already been applied: To prevent email spamming, in the early 1990s Dwork and Naor~\cite{Dwork} suggested using squaring roots over finite fields puzzles as functions which take a predetermined time to compute, and are straightforward to verify. However, their work was considered impractical because one has to use rather large finite fields to make the algorithm useful, and the libraries for handling multiple-precision arithmetic at the time of the suggestion of the algorithm were orders of magnitude slower than current ones.
 
Based on~\cite{Dwork}, Boneh \textit{et al.} designed VDFs as functions that can be evaluated in a given amount of sequential steps and verified in an exponentially shorter time~\cite{Boneh2019}. The main innovation was to propose a setup phase to set a trusted environment allowing the functions to be universally verifiable. This trusted environment sets the public parameters of the VDF, including its difficulty which determines the amount of time spent on computing. Any node who needs to solve the VDF will use the public parameters to perform certain sequential computations. Some VDFs also allow generating a \textit{proof} to facilitate the verification from the other participants. This creates a set of three algorithms, \textit{computation}, \textit{proof}, and \textit{verification}, which are formally described in~\cite{Boneh2019}.

To date, the main VDF constructions are based on modular exponentiations, where Pietrzak~\cite{Pietrzak} and Wesolowski~\cite{Wesolowski} suggest to iteratively compute $\tau$ squarings in an RSA group, with $\tau$ large. Modular exponentiation is a deeply studied computational problem due to its straightforward importance in many critical public key encryption algorithms. The best proven lower complexity bound remains valid even assuming unbounded parallelism~\cite{Borodin1992, Dimitrov1998}. For this reason, it is possible to obtain extremely accurate assessments about the timing of the modular exponentiation operations (and the corresponding VDF characteristics) based on the best available ASIC designs for these particular operations. On the contrary, other VDFs, e.g., the ones based on pairing over elliptic curves~\cite{DeFeo}, have been subject to much fewer research studies and its parallel complexity remains unknown. 

\section{System model and problem statement}\label{sec:model}

We formalize here the system model inspired by the IOTA network~\cite{Popov2018}. We assume a network $\mathcal{N}$ with $n$ nodes. A node can be either standard (e.g., IoT device, smartphone, laptop) or specialized (e.g., FPGA, ASIC) hardware, and is connected to $m\ll n$ neighbors. Every node is involved in the generation and the verification of transactions, which transfer tokens between two nodes, or simply carry data. While the verification task is fundamental to reach consensus in the network, in this work we are interested in the transaction generation which can be exploited to target specific nodes through DoS attacks. As this transaction generation is not limited by any fee, without an appropriate access control mechanism, a node could theoretically generate an infinite number of transactions per second, leading to network disruption.

We assume that a node has to evaluate a function $f$ before issuing a transaction. Let the value $\theta_i(f)$ represent the throughput at node $i$ when evaluating function $f$. For instance, assume that $f$ is SHA-256, the hash function used in Bitcoin: The ASIC Ebit E10 can compute 18,000,000 MHash/s ($\theta_{ASIC} = 1.8\times 10^{13}$), while a standard CPU i7 3930K can only compute 66 MHash/s ($\theta_{CPU} = 6.6\times 10^7$). Depending on the function used, the unit of measurement may change. The goal of our work is described in the following:

\begin{mdframed}[backgroundcolor=gray!20]
\textit{Problem statement.} Choose a function $f$ such that the maximum \textit{speedup} in throughput $S$ is minimized, i.e.,
\begin{equation}\label{eq:problem}
    \arg\min_{f} \ S(f)\triangleq \frac{\max_{i\in\mathcal{N}}\theta_i(f)}{\min_{i \in\mathcal{N}}\theta_i(f)}.
\end{equation}
\end{mdframed}

This task is particularly challenging in heterogeneous networks, where devices with very different capabilities are present. Consider, for example, a DoS prevention mechanism based on the solution of a PoW: Specifically, before issuing a transaction each node is required to find an input for a hash function which output begins with a certain amount of zeros in its binary representation. The only known way of finding such an input is trying randomly the inputs until finding a suitable output. The more zeros are required, the more difficult the PoW is. However, if the PoW is too easy, then an ASIC can speed up the solution of the puzzle by several orders of magnitude and outperform non-specialized hardware; on the other hand, if the PoW is too difficult, then DoS attacks are prevented, but low power devices will not have enough computational power to issue transactions in a reasonable time. In the following section, we propose a DoS protection mechanism where the function $f$ is a VDF.
\section{Protocol design}\label{sec:protocol}

\begin{figure}
\centering
\includegraphics[width=\linewidth]{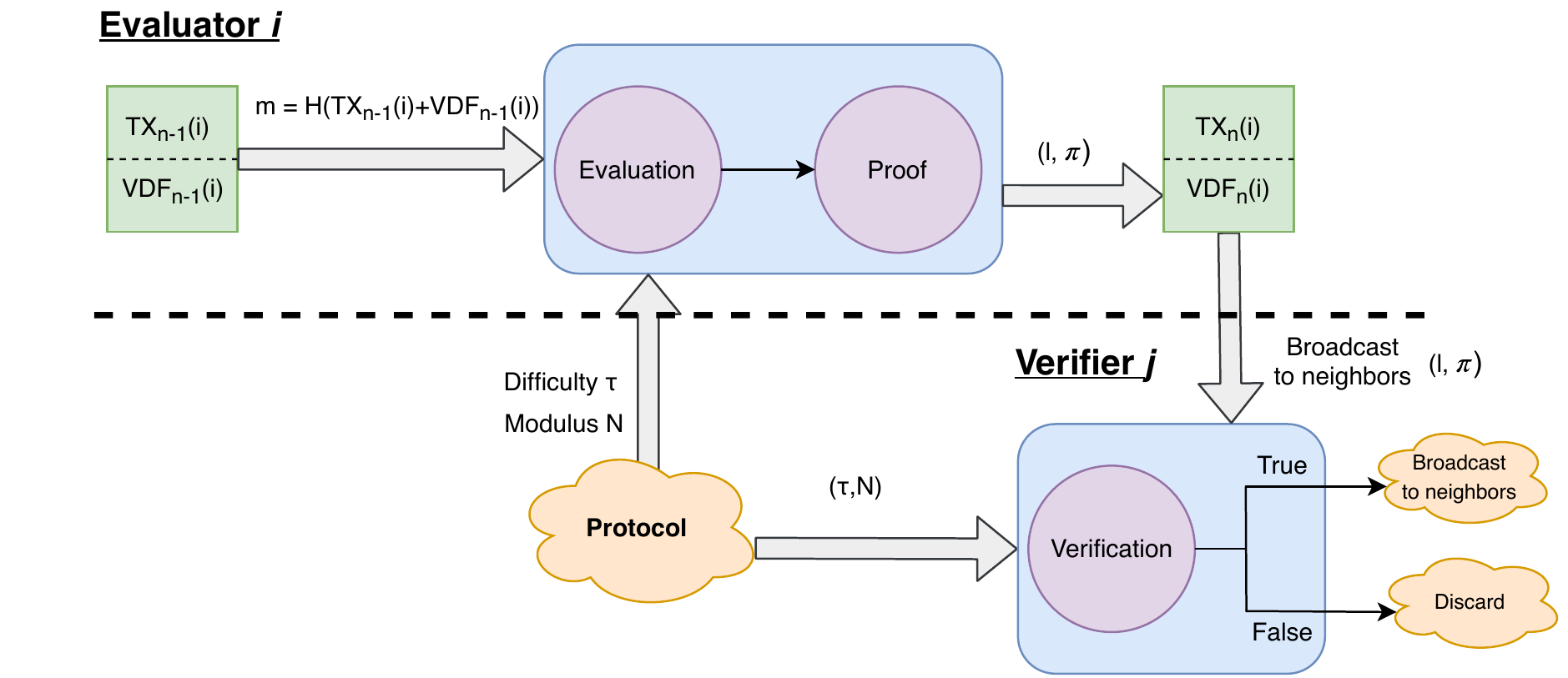}
\caption{VDF integration in the IOTA protocol: The evaluator node uses its last transaction as the input for the new VDF (above); a verifier node receives the transaction and the proof from node $i$, and verifies the correctness of the VDF (below). The \textit{protocol} cloud represents the values known by all nodes.}
\label{fig:vdf_chain}
\end{figure}

In this section, we present our DoS prevention mechanism (Figure~\ref{fig:vdf_chain}) which is defined by the following algorithms:
\begin{itemize}[leftmargin=*]
    \item \textit{Evaluation.} When node $i$ decides to generate transaction $n$, it is required to solve a VDF such that its input is the hash of transaction $n-1$ issued by the same node.
    \item \textit{Proof.} Node $i$ also generates a proof to facilitate the verification task, which gossips along with the transaction.
    \item \textit{Verification.} When a new transaction is received, node $j$ verifies whether the VDF has been solved correctly. If yes, it forwards the transaction (and the proof) to its neighbors, or discards it otherwise.
\end{itemize}

All nodes share the following public inputs\footnote{Depending on the degree of decentralization targeted, these inputs can either be preset by the network manager or generated in a distributed way by the nodes.}:
\begin{itemize}[leftmargin=*]
    \item \textit{VDF difficulty.} A difficulty $\tau\in\mathbb{N}$, which indicates the number of sequential operations to solve. This difficulty can be adapted according to, e.g., node's reputation to mitigate Sybil attacks.
    \item \textit{RSA modulus.} A modulus $N=p\cdot q$ which has bit-length $\lambda$ (typically 1024, 2048, or 3072), and is the product of two prime numbers of the same order. Due to the similarity with the Rivest-Shamir-Adleman (RSA) cryptosystem~\cite{Rivest1978}, we refer to $N$ as RSA modulus. The security of the entire mechanism relies on the length of $N$: the longer is the modulus, the more difficult is to find its factorization. For further information about RSA modulus factorization, we refer the interested reader to~\cite{Cybernetica2016}. How to securely generate the RSA modulus without any sort of centralization is an active research topic~\cite{Boneh1997,Frederiksen,Damgard2010}. %Since this is not the main focus of the work, in this paper we assume the modulus is known by all nodes while its factorization remains unknown.
    \item \textit{Cryptographic hash function.} A hash function $H(m)$ such that $H(m): \{0,1\}^* \mapsto [0,2k]$, where $m$ is a binary message, and $k$ is a security parameter. %We require $2k$ bits as an input to guarantee a $k$ bit-level security against preimage attacks \cite{Rogaway2009}. In other words, breaking the hash function's security would take approximately $2^k$ computations. For a hashing function with $2k$ bit-long output, it has a pre-image resistance of $2k$ bits and a collision resistance of $k$ bits, hence the output size of $H$. Pre-image resistance means it is practically impossible to guess an input for a given output and collision resistance means that finding two inputs yielding the same output is hard.
    A typical value for $k$ is 128, and one could use SHA-256.
    Having this in mind, we also define $H_{prime}(m)$ which gives the smallest prime greater or equal to $H(m)$ for a given message $m$:
        \begin{equation*}
            H_{prime}(m) = \min\{x \in\mathbb{N} \ | \ x \geq H(m) \wedge x \ \texttt{prime}\}.
        \end{equation*}
    This hash function will be used in the evaluation and the proof of the VDF.
    \end{itemize}

%A fundamental question is how to generate the RSA modulus without disclosing its factorization to any of the network participants or relying on any sort of centralization. There are several existing solutions on distributed RSA keys generation~\cite{Boneh1997, Damgard2010, Hazay, Gilboa, Dottax, Frederiksen}. The main idea is to run a multi-party computation where each party generates two random numbers $p_i$, $q_i$ of size $\lambda/2$. The algorithm runs distributed trial divisions over small prime numbers on both $\sum\limits_ip_i$ and $\sum\limits_iq_i$ to easily discriminate wrong (i.e., non-prime) candidates. After that, it runs a distributed multiplication such that $N=\sum\limits_ip_i\cdot\sum\limits_iq_i$, and computes a biprimality test. In our protocol, we impose the RSA modulus to be recomputed regularly to decrease the chances of being inferred somehow by an attacker.

Our VDF design also addresses resource constraints in terms of (i) bandwidth and (ii) computational capabilities: (i) Little overhead (in bytes) should be added to the transaction as increasing the transaction size would lead to a lower throughput; (ii) any device must be able to verify the correctness of the VDF solution in short time. We choose to build our mechanism on top of the Wesolowski construction~\cite{Wesolowski} as it currently provides the best tradeoff between verification time and lightness of the outputs~\cite{Attias2020}. We rigorously describe the VDF evaluation and proof in Section~\ref{sec:wesolowski}, the VDF verification in Section~\ref{sec:verification}, and we provide a computational complexity analysis in Section~\ref{sec:complexity}.

\subsection{Evaluation and proof generation} \label{sec:wesolowski}
Each time a node decides to issue a transaction, it has to evaluate a VDF. The evaluation takes as an input a binary message $m\in\{0,1\}^*$, which we enforce to be the previous transaction issued by the same node\footnote{In the case the node is issuing its first transaction, it will use the timestamp included inside the transaction.}. In this way, the protocol ensures the sequentiality of the VDFs evaluations because a node has to wait for a VDF to complete before starting a new evaluation. In other terms, it is not possible to evaluate several VDFs in parallel to overcome node's allowed throughput. The challenge for the solver is computing $y$ such that
\begin{equation}\label{eq:eval}
    y = x^{2^\tau}\Mod N,
\end{equation}
with $x\triangleq H(m)$. This can be done only using $\tau$ sequential squarings. We stress that it is crucial that the factorization of $N$ remain unknown, otherwise the computation of Eq.~\eqref{eq:eval} could be reduced to extremely simpler modular exponentiations.

Unlike PoW, in VDFs there is no trivial way to verify whether a node has correctly solved Eq.~\eqref{eq:eval}. To favor this task, we impose that the node computes the following numbers:
\begin{itemize}[leftmargin=*]
    \item $l=H_{prime}(x+y)$, that is the smallest prime number larger than the hash of the sum of the VDF input $x$ and output $y$;
    \item $\pi=x^{\floor{2^\tau/l}}$, that is an exponentiation of $x$ by the result of a long division of $2^\tau$ by $l$.
\end{itemize}
The resulting \textit{proof} is the pair $\{l, \pi\}$, which has a size of $\lambda+2 k$ bits. The summation between $x$ and $y$ ensures that the challenge has been successfully solved. Furthermore, to reduce transaction size, we can avoid sending the output $y$: In fact, a verifier can easily recompute $y$ from the knowledge of $l$ and $\pi$. We will elaborate on this idea in the next paragraph. Algorithm~\ref{alg:eval_proof} summarizes the evaluation and the proof.

\begin{algorithm}[ht]
\SetAlgoLined
\SetKwInOut{Input}{input}
\SetKwInOut{Output}{output}
\Input{$m\in\{0,1\}^*$, $\tau\in\mathbb{N}$}
\Output{$\pi\in[0,N-1]$, $l$ prime $\in [0,2^{2k}-1]$}

 $x\leftarrow H(m)$\\
 $y\leftarrow h$\\
 \For{$k \leftarrow 1$ to $\tau$}{
  $y\leftarrow y^2$
 }
 
 $l\leftarrow H_{prime}(x+y)$\\
 $\pi=x^{\floor{2^\tau/l}}$\\
 \textbf{return} $(\pi,l)$
 \caption{Evaluation and proof of the VDF}\label{alg:eval_proof}
\end{algorithm}

%\vspace{-6mm}

\subsection{Verification}\label{sec:verification}
In the previous subsection, we have introduced the proof necessary to verify the VDF solution in short time. The first step is to reconstruct the VDF output $y$ using
\begin{equation}\label{eq:verif}
    y=(\pi^l\cdot x^r) \ \Mod N,
\end{equation}
where $r=2^\tau\ \Mod l$. To verify that the solver has correctly computed $y$, the verifier will simply check if $H_{prime}(x+y)=l$. The verification task is described in Algorithm~\ref{alg:verif}.

As an additional remark, the multi-exponentiation part of the verification phase given by Eq.~\eqref{eq:verif} takes the vast majority of the total verification time. Hence, optimizing this computation can be crucial. In our simulations, we propose a way to speed up this task through a multi-exponentiation technique based on the algorithm presented in~\cite{Yen1994}.

\begin{algorithm}[ht]
\SetAlgoLined
\SetKwInOut{Input}{input}
\SetKwInOut{Output}{output}
\Input{$x,\tau,\pi,l$}
\Output{$True$ or $False$}

 $x\leftarrow H(m)$\\
 $r\leftarrow 2^\tau\ \Mod l$\\
 $y\leftarrow \pi^l\cdot x^r\Mod N$\\
 
 \uIf{$l=H_{prime}(x+y)$}{
  \textbf{return} $True$
 }
 \Else{
  \textbf{return} $False$
 }
 \caption{Verification of the VDF}\label{alg:verif}
\end{algorithm}
%\vspace{-6mm}

\subsection{Computational complexity analysis}\label{sec:complexity}

The computational complexity of the proposed protocol depends on the time complexities needed to evaluate $y$ both by the solver through Eq.~\eqref{eq:eval}, and by the verifier through Eq.~\eqref{eq:verif}.

\subsubsection{Computational complexity of the evaluation}

The computation of Eq.~\eqref{eq:eval} consists of a single computational primitive, i.e., modular squaring. The most important thing is that the computational process is strictly sequential, that is every new modular squaring operation can start only when the previous squaring operation is completed.

\begin{result}\label{th:sequential}
    The solver of Eq.~\eqref{eq:eval} is required to \textit{exactly} solve $\tau$ modular squarings.
\end{result}

To date, there is no known algorithm capable to guarantee parallel execution of the modular squaring function. Although there are few chances to find a rigorous proof that this function does not admit parallel execution, the majority of cryptographers accept this conjecture as correct~\cite{VDFResearch}. 
%The state of computational complexity theory is such that there is absolutely no hope to find a strict mathematical proof that this function does not admit parallel execution. The majority of cryptographers accept this conjecture as correct~\cite{VDFResearch}.
Indeed, in this field many results are based on conjectures that seem plausible: For instance, this is the case of RSA, based on the conjecture that the factorization problem is computationally intractable; for large RSA key sizes, no efficient method for solving this problem is known.

\subsubsection{Computational complexity of the verification}

The computation of Eq.~\eqref{eq:verif} can be accomplished in various ways. It is important to point out that the two modular exponentiations \emph{do not have} to be computed separately. The easiest way to see why this is the case is to consider the computation of $z = x^2\cdot y^2$. If one computes separately $x^2$, $y^2$, and their product, one would need three multiplications, whereas representing $z$ as $(x\cdot y)^2$ needs only two multiplication steps.

We have the following result:

\begin{result}\label{th:verif_result}
    The number of operations for the verifier to compute Eq.~\eqref{eq:verif} requires \textit{at most}
    \begin{equation}\label{eq:verif_bound}
    \mathcal{O}\left(\frac{2k}{\log_22k}\right)
    \end{equation}
    multiplications over $\lambda$-bit numbers.
\end{result}

\begin{IEEEproof}
The best algorithms for evaluating $y = \pi^l\cdot x^r$ are based on finding very short vector addition chains for the vector $\{l,r\}$. It is well known that the length of the vector addition chain is asymptotically equal to the length of the scalar addition chain for the largest component of the vector. Therefore, the number of operations for the verifier to compute Eq.~\eqref{eq:verif} is equal to
    \begin{equation*}
        \log_2(\max\{l,r\}) + \mathcal{O}\left(\frac{\log_2(\max\{l,r\})}{\log_2(\log_2(\max\{l,r\})}\right)
    \end{equation*}
    multiplications \cite{Pippenger1980}. Note that $\max\{l, r\} \leq 2^{2k}$ since $r < l$ by definition, and $l$ is at most $2^{2k}$. Then, after simple calculations, one can get obtain the bound of Eq.~\eqref{eq:verif_bound}.
\end{IEEEproof}

From the above, we can deduct the following result:

\begin{result}
    The time needed to verify the correctness of the VDF output is independent of its difficulty $\tau$.
\label{result:time_independant}
\end{result}

The previous results are important as they can provide a clear understanding of the physical limits of the verification time, which is directly linked to the potential maximum number of transactions per second allowed in the network. We will elaborate more on this in the next section, where we perform experimental simulations.

%%%%%%%%%%%%%%%%%%%%%%%%%%% Merge with the previous text

%The efficiency of the verification lies in the fact that the evaluator only has to compute two exponentiations with exponents of size at most $2\cdot k$ and radix size at most $\lambda$ and then finding the next prime number larger than $H(x+y)$. This leads to the following result:
%\begin{result}
%\label{theo:time_verif}
%Assume one has the pair $\{l, \pi\}$, as described in this section. Hence, the time needed to verify whether the corresponding VDF has been correctly computed is a function of the security parameters $\lambda$ and $k$, and independent of the difficulty $\tau$.
%\end{result}

%\begin{proof}
%The first step in the verification is to calculate $r$, which can be efficiently computed knowing that $l$ is a prime number which yields that $\phi(l) = (l-1)$ and then $2^\tau\Mod l = 2^{\tau\Mod \phi(l)}\Mod l$.

%Then, the computation of $y$ requires the modular exponentiations of $\pi^l$ and $x^r$. In these two exponentiations, the radix is at maximum a $\lambda$ bits number and the exponent a $2k$ bits number.

%Finally, the $H_{prime}$ function can be computed in constant time.

%In conclusion, the verification complexity is only a function of $\lambda$ and $k$ and is, therefore, independent of $\tau$.
%\end{proof}
\section{Simulation results}\label{sec:simulation}

In this section, we evaluate the performance of our VDF-based DoS prevention mechanism. The algorithms related to the VDF, namely Algorithms~\ref{alg:eval_proof}-\ref{alg:verif}, have been implemented in C++ using the \texttt{GMP} library for multi-precision modular arithmetic. The code is executed on a laptop using an Intel i7-7820HQ @ 2.90GHz (\textit{CPU}) and on a Raspberry Pi 3 Model B (\textit{IoT}). As for completeness, we consider the most up-to-date FPGA solving VDFs according to the results of the \textit{VDF Alliance FPGA contest}~\cite{VDFAlliance}. %As an input message, we consider ???
Unless otherwise stated, in the simulations we use a modulus size $\lambda=2048$, and a security parameter $k=128$.

\begin{figure}
\centering
\includegraphics[width=0.9\linewidth]{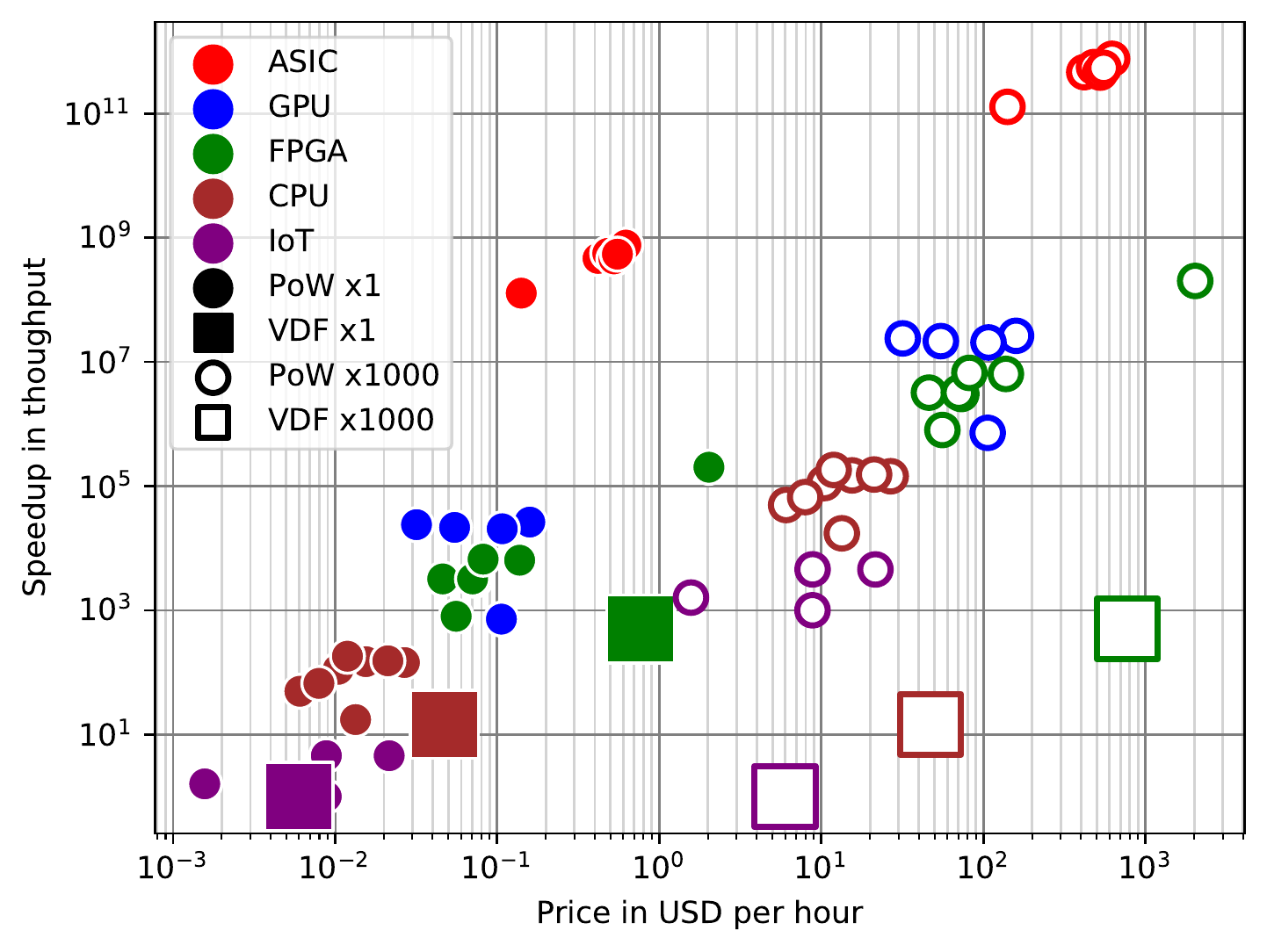}
\caption{Speedup for different devices as a function of their cost (hardware plus electricity consumption) in USD per hour to compute VDF or PoW. We also show the speedup of combining 1000 devices in a pool (empty markers). The baseline is the weakest device for VDF and PoW.}
\label{fig:performance}
\end{figure}

Recall the problem statement that requires to find a function $f$ to minimize the maximum speedup in throughput. In Figure~\ref{fig:performance} we compare two functions, our VDF and a PoW based on SHA-256. Specifically, we display the speedup in throughput for different devices computing either VDF or PoW as a function of their estimated price per hour, which is computed by amortizing the hardware flat price over one year added to the power consumption (we take the electricity price in China as a reference). Data related to PoW are taken from~\cite{BitcoinWiki}. The slowest devices are taken as a baseline.

From the plot, we can clearly see that specialized hardware is able to exploit PoW parallelizability, which leads to a dramatic speedup. On the other hand, VDF bounds the possible speedup to less than three orders of magnitude: While it is not possible to reduce or parallelize the sequential steps of VDF, an FPGA can actually exploit optimized arithmetic operations within each single modular squaring. We believe that the speedup for VDF can be considered a worst-case scenario, and other IoT devices may show significant better results: The reason why \textit{existing} VDFs do not offer a good performance on this kind of IoT is the fact that they all use a large number of multi-word divisions. On a Raspberry Pi, these divisions have to be entirely implemented at software level, which slows the execution time many times. Although these considerations do not apply to SHA-256, which requires very basic operations such as bit shift or multiplications, the effectiveness of our solution remains clear. We have also performed simulations with other hashing functions, and they show similar outcomes.

We additionally consider pools of 1000 devices of the same type, and we display them in the same figure through empty markers. As PoW is linearly paralellizable, the speedup increases by a factor of 1000x when using 1000 machines; conversely, hardware farms cannot contribute to increase the speedup for VDFs, where only the estimated cost goes up. This result alone should justify the choice of using VDFs as a DoS prevention mechanism in IoT networks.

%In Figure~\ref{fig:energy}, we computed the energy efficiency of the different hardware. This is measured in Hashes per Joule or Squarings per Joule and represents the computing power obtained when spending 1 Joule in electrical energy. This is used by dividing the hashrate or squaring rate by the electrical power of the hardware. We have displayed here only VDF and SHA-256 for IoT, CPU, and FPGA and not ASIC, due to the lack of data for the missing configurations. Again we can see that using specialized hardware has much less interest for VDF than PoW, and as the electrical consumption is an important economical aspect of the cryptocurrencies, it reduces the economical value of spamming the network with specialized hardware.

%\begin{figure}
%\centering
%\includegraphics[width=\linewidth]{images/energy_factor.pdf}
%\caption{Comparison of verification time using a naive implementation and Lenstra's algorithm with $w=2$ with variation of $k$ and $\lambda$.}
%\label{fig:energy}
%\end{figure}

\begin{figure}
\centering
\includegraphics[width=0.9\linewidth]{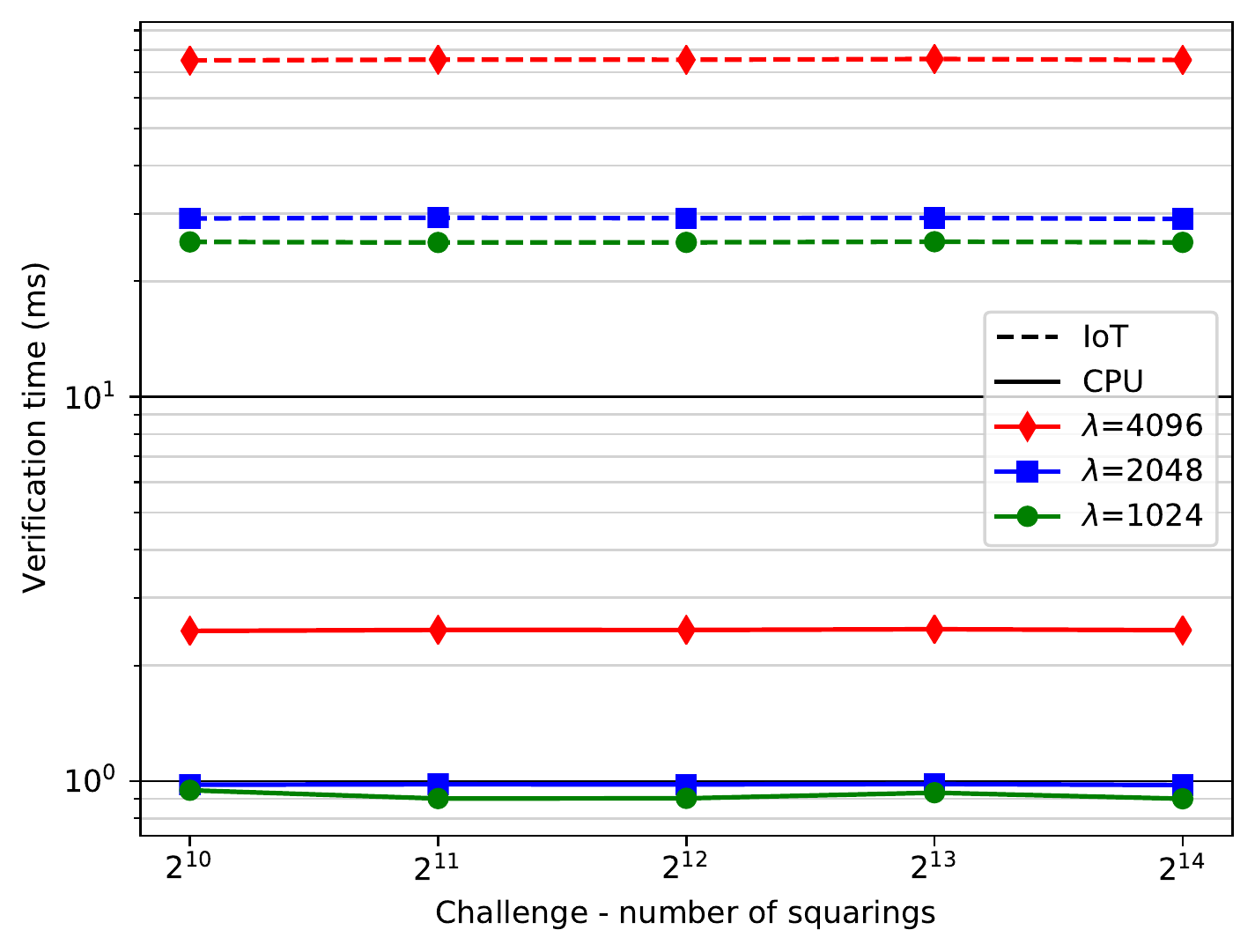}
\caption{Verification time as a function of the modular squarings $\tau$ for different modulus size $\lambda$ ($k=128$).}
\label{fig:verification_time}
\end{figure}

Figure~\ref{fig:verification_time} shows the verification time of the VDF in case of IoT and CPU. The x-axis represents the challenge $\tau\in\{2^{10}; 2^{11}; 2^{12}; 2^{13}; 2^{14}\}$, i.e., the number of squarings needed to evaluate the VDF, and the y-axis shows the time spent in the verification task in \textit{ms}. The figure also shows the impact of the modulus length $\lambda\in\{1024; 2048; 4096\}$ on the verification time. The plot validates \textit{Result 3}, as the verification time is indeed independent of the difficulty $\tau$. The performance speed up between CPU and IoT shows a factor of 20 in terms of verification time, being between 1 and 3 ms for the former and between 25 to 75 ms for the latter. The limited resources of IoT devices are confirmed once again to be an important aspect to consider in the design of our DoS prevention mechanism.

\begin{figure}
\centering
\includegraphics[width=0.9\linewidth]{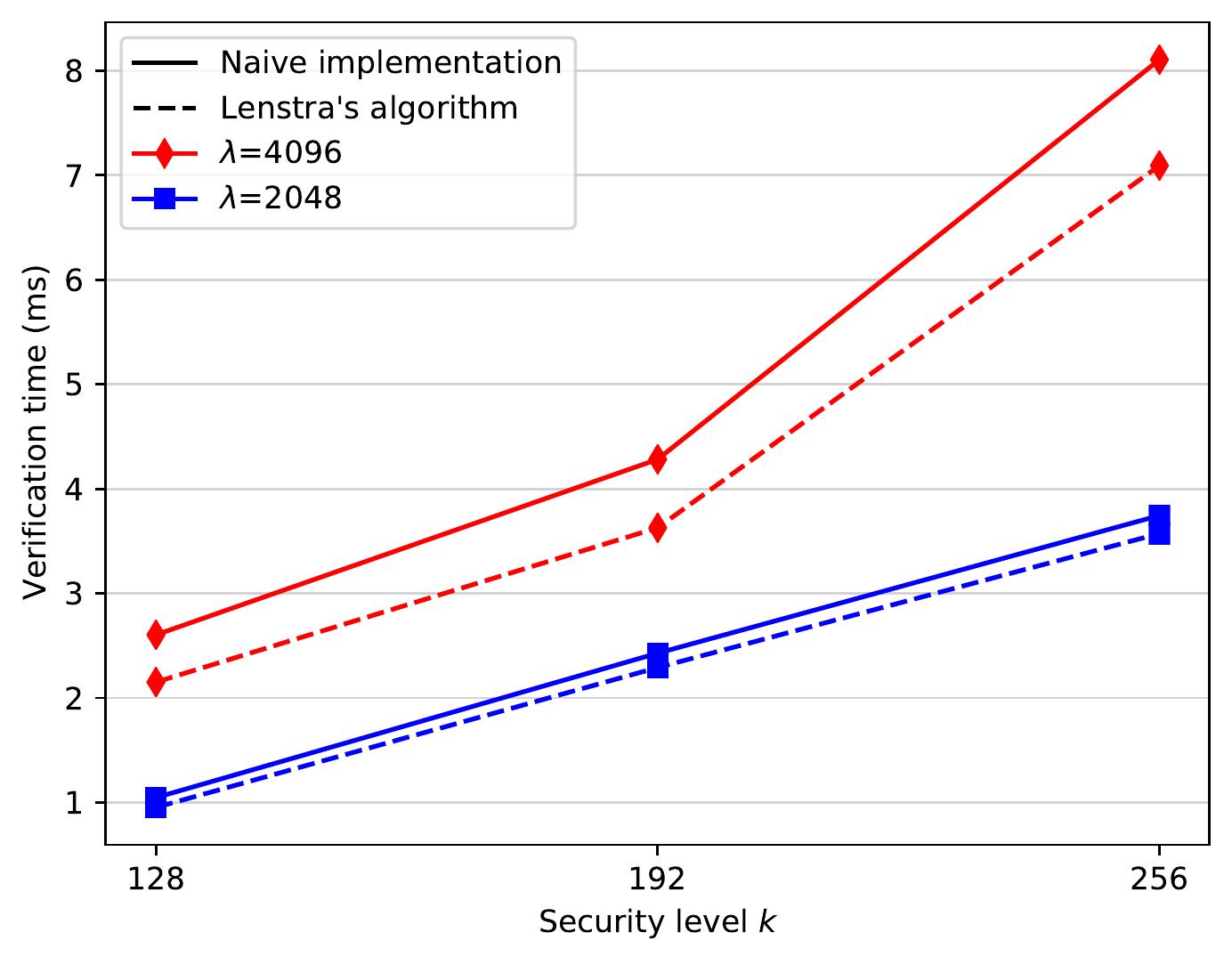}
\caption{Verification time using multi-exponentiation techniques for different security levels $k$ and modulus sizes $\lambda$ on a CPU.}
\label{fig:multi_exp}
\end{figure}

In Figure~\ref{fig:multi_exp}, we show the verification time for laptop using different multi-exponentiation techniques to compute Eq.~\eqref{eq:verif}: \textit{Naive implementation} computes $\pi^l$ and $x^r$ separately and then multiplies each term; \textit{Lenstra's algorithm} is based on the technique described in~\cite{Yen1994, Moller}, and we use a window-scanning size of 2 bits. We compare three security level $k$ ($128$, $192$ or $256$ bits) and two modulus length $\lambda$ ($2048$ and $4096$). From the figure, we can observe that the Lenstra's algorithm performs between 12\% to 18\% better than the naive implementation when using a 4096 bits modulus and between 4\% to 8\% when using a 2048 bit modulus. One of the most promising possibilities for further improving the verification time is to use some of the currently proposed parallel algorithms for multi-exponentiations~\cite{Borges2017}. This is a topic of an on-going research within our group.

\section{Conclusion}\label{sec:conclusion}
In this paper we have proposed the usage of a VDF to mitigate DoS attacks in IoT-oriented DLTs. Throughout the paper, we have formulated a rate limiting protocol based on a modular exponentiation VDF, which introduces low overhead in the network; additionally, we have provided some fundamental analytical findings, and we have validated our design through actual implementation. We stress that, since VDFs are a new field of research, which has been mostly studied from a theoretical point of view, little to no experimental results are available to date, and our paper is a pioneer in this domain.

Due to the novelty of the VDFs, many are the possible future directions. To remain in the scope of the paper, we can mention a few: Software optimization to efficiently solve VDFs on IoT devices; development of VDFs that do not employ multi-word division operations; tradeoff analysis between nodes computational capabilities and energy consumption; optimization of the verification time through multi-exponentiation and parallelization.

\bibliographystyle{IEEEtran}
\bibliography{IEEEabrv,bibliography}

%\begin{thebibliography}{00}
%\bibitem{b1} G. Eason, B. Noble, and I. N. Sneddon, ``On certain integrals of Lipschitz-Hankel type involving products of Bessel functions,'' Phil. Trans. Roy. Soc. London, vol. A247, pp. 529--551, April 1955.
%\end{thebibliography}

\end{document}